\documentclass[aps,pra,groupedaddress, superscriptaddress, amsfonts,twoside,twocolumn,amssymb,final]{revtex4}

\usepackage{amsmath}
\usepackage{amsfonts}
\usepackage{amssymb}
\usepackage{graphicx}

\def\ketbra#1#2{\vert#1\rangle\langle#2\vert}

\def\ket#1{\vert#1\rangle}
\def\bra#1{\langle#1\vert}

\DeclareMathOperator{\id}{id}

\def\EE{{\cal E}}\def\RR{{\cal R}}\def\NN{{\cal N}}

\begin{document}
\title{Comment on ``Optimum Quantum Error Recovery using Semidefinite Programming''}
\author{M. Reimpell}
\author{R.~F. Werner}
\affiliation{Inst. Math. Phys., TU-Braunschweig, Mendelssohnstra{\ss}e 3,
  D-38106 Braunschweig, Germany}
\author{K. Audenaert}
\affiliation{Department of Physics, Blackett Laboratory, Imperial College London,
Prince Consort Road, London SW7 2BW, UK}

\begin{abstract}
In a recent paper ([1]=quant-ph/0606035) it is shown how the optimal
recovery operation in an error correction scheme can be considered
as a semidefinite program. As a possible future improvement it is
noted that still better error correction might be obtained by
optimizing the encoding as well. In this note we present the result
of such an improvement, specifically for the four-bit correction of
an amplitude damping channel considered in [1]. We get a strict
improvement for almost all values of the damping parameter. The
method (and the computer code) is taken from our earlier study of
such correction schemes (quant-ph/0307138).
\end{abstract}

\maketitle

\section{Introduction}
In a recent paper Fletcher, Shor and Win \cite{lit:shor} analyze
error correction schemes obtained by ab initio optimization, rather
than the adaptation of classical coding techniques. The basic idea
\cite{lit:iter,talk:a2,talk:dpg,poster:cam,lit:phd} is that both
encoding $\EE$ and recovery (or decoding) $\RR$ can be arbitrary
channels, and that for a given number $n$ of invocations of a noisy
discrete memoryless channel the objective is to bring the channel
$\EE\NN^{\otimes n}\RR$ as close to the identity as possible. If the
so-called channel fidelity used as a figure of merit (as in
\cite{lit:shor} and \cite{lit:iter}) the optimization is clearly a
semidefinite problem, by virtue of Jamiolkowski-Choi duality.

In \cite{lit:iter} we presented an alternative algorithm for this
semidefinite problem, and used it to generate optimal codes for
various noisy channels, by alternatingly optimizing the encoding
channel and the recovery channel. In \cite{lit:shor} only the
recovery is optimized, which already gives a marked improvement some
over previously known codes, specifically for the case of the
four-bit correction of the amplitude damping channel \cite{lit:leung}.
The possibility of further improvements by optimizing also the
encoding is noted in the discussion (citing also \cite{lit:iter}).

It so turns out that the required computation (even for the same
test case) was already done in the autumn of 2003 in a collaboration
between the first two authors and the third author of this note,
with the aim of checking the power-iteration method of
\cite{lit:iter} against the better established semidefinite method.
Since these results directly support the perspective forwarded in
\cite{lit:shor}, we felt it appropriate to make them immediately
available.

\section{Results}

For the theoretical background we refer to either \cite{lit:iter} or
\cite{lit:shor}. In both papers similar ideas and notations are
used, so they should be readily accessible from each other. The {\em
amplitude damping channel} is the qubit channel $\NN=\NN_\gamma$
with Kraus operators
\begin{equation}\label{kraus}
    K_0=\left(\begin{array}{cc}1&0\\0&\sqrt{1-\gamma}\end{array}\right)
    \quad\mbox{and}\quad
    K_1=\left(\begin{array}{cc}0&\sqrt{\gamma}\\0&0\end{array}\right).
\end{equation}
These channels form a semigroup
($\NN_\gamma\NN_\varepsilon=\NN_{\gamma\varepsilon})$, which
contracts to the first ``unexcited'' basis state. The channel
fidelity of a noisy channel $\NN$ is defined as
\begin{equation}\label{fides}
    F(\NN):=\bra\Omega(\NN\otimes\id)(\ketbra\Omega\Omega)\ket\Omega,
\end{equation}
where $\Omega$ is a maximally entangled vector. Note that this
requires input and output of the channel to be systems with the same
Hilbert space, which is adapted to comparing the channel with the
identity, which is the unique channel with $F(\NN)=1$. This is also
closely related \cite{lit:horodecki} to the average fidelity for pure
input states (with the average taken according to the unitarily
invariant measure). The main virtue of choosing this fidelity as a
figure characterizing the deviation from the identity is that it is
linear in $\NN$. Such a linear criterion is possible only because
the ideal channel is on the boundary of the set of channels.

\begin{figure}
  \begin{center}
    \includegraphics*[width=7cm]{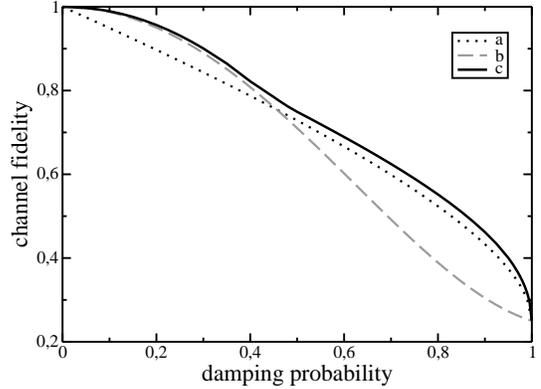}
    \caption{\label{fig:seessaw}Error correction results for the amplitude damping channel (four copies) using three methods: (a) no coding, (b) optimized decoding by Fletcher et al.\ \cite{lit:shor} (with encoding by Leung et al. \cite{lit:leung}) and (c) our iterative optimization of both, encoding and decoding \cite{lit:iter}.}
  \end{center}
\end{figure}%

The curve $\gamma\mapsto F(\NN_\gamma)$ is the dotted line in
Fig.~1. The other lines represent $\gamma\mapsto
F(\EE_\gamma\NN_\gamma^{\otimes4}\RR_\gamma)$ for various choices of
$\EE_\gamma$ and $\RR_\gamma$. The dashed line uses the encoding $\EE_\gamma$ by Leung et al. \cite{lit:leung}
\begin{align*}
\ket{0}_L & = \frac{1}{\sqrt{2}}(\ket{0000} + \ket{1111}), \\
\ket{1}_L & = \frac{1}{\sqrt{2}}(\ket{0011} + \ket{1100}),
\end{align*}
with an optimized decoding $\RR_\gamma$ \cite{lit:shor}. In fact, we computed this line by our routines, and it coincides to within pixel resolution with the graph in \cite{lit:shor}. The solid line is the result of the  iteration \cite{lit:iter}, in which  $\EE_\gamma$ and $\RR_\gamma$ are alternatingly optimized, keeping the other operation fixed. This iteration has a strict improvement over the Leung code for $0 < \gamma < 1$.

\section{Discussion}
The methods in \cite{lit:shor} and \cite{lit:iter} have the following characteristic features:
\begin{itemize}
 \item For a known channel these methods yield excellent results, without using any special properties of the channel like symmetry etc.. 
\item The optimization of either encoding or decoding is a semidefinite problem for which the solution is a certified global optimum. The process of alternatingly optimizing these therefore improves the objective in every step, and hence converges to a local optimum. However, there is no guarantee for having found the global optimum.
\item  The methods suffer from the familiar explosion of difficulty as the system size is increased. Correction schemes like the five qubit code can still be handled on a PC, but a nine qubit code would involve optimization over $2^{10}\times2^{10}$-matrices, which is set up by multiplying and contracting some matrices in $2^{18}$ dimensions. This may be possible on large machines, but it is clear that these methods are useless for asymptotic questions.
\item The iteration method replacing the semidefinite package in \cite{lit:iter} has a slight advantage here, because it works with a fixed number of Kraus operators. So for the encoding one can put in by hand an isometric encoding, which, as our study shows, is often optimal. This cuts down on dimensions, at least for the optimization of encodings.
\item For asymptotic coding theory one still needs codes, which can be described also for very large dimensions, be it by explicit parameterization or by a characterization of typical instances of random codes. It is here that methods transferred from classical coding theory will continue to be useful. 
\end{itemize}


\end{document}